\documentclass[11pt,twoside]{article}

\usepackage{asp2006}
\usepackage{epsf}
\usepackage{psfig}
\usepackage{lscape}

\begin{document}
\title{Methods for determining AGB mass loss rates based on radio data}   %%% Fill in title
\author{Fredrik L. Sch{\"o}ier}
                %%% Fill in author names
%\affil{}    %%% Fill in author affiliations
\affil{Stockholm Observatory, AlbaNova University Center, SE-10691, Sweden}

\begin{abstract} %%% Abstract to run on from here.
In the radio regime the mass-loss rate of AGB stars is best probed using molecular (and atomic) line emission arising in the CSE formed by the stellar wind. The numerical modelling of the circumstellar emission where intricate interplays between physical and chemical processes take place, is a challenge. The derived mass-loss rates depend crucially on the assumptions in the circumstellar model, of which some can be
constrained if enough observational data exist.  Therefore, a reliable
mass-loss-rate determination for an individual star requires, in
addition to a detailed radiative transfer analysis, good observational
constraints in the form of multi-line observations and radial
brightness distributions. Of the methods used to estimate mass-loss rates from galactic AGB stars those based on radiative transfer modelling of CO line emission are most commonly used and possibly also the most accurate. Typically, CO multi-transitional observations can constrain the mass-loss rate to better than 50\%, within the adopted circumstellar model. Comparison with complementary methods, such as estimates based on dust radiative transfer modelling coupled with a dynamical model, are consistent within a factor of three.

\end{abstract}

%%% MAIN BODY OF TEXT GOES HERE. CONSULT "INSTRUCTIONS FOR AUTHORS USING
%%% LATEX2E MARKUP", SECTIONS 2.3-2.6 FOR HELP WITH EQUATIONS, FIGURES,
%%% AND TABLES.

%\section{}   %%% Top level section head (remove "%" symbol)
%\subsection{}   %%% Second level section head (remove "%" symbol)
%\subsubsection{}   %%% Lowest level section head (remove "%" symbol)
%\section*{}    %%% Unnumbered top level section head (remove "%" symbol)
%\subsection*{}   %%% Unnumbered second level section head (remove "%" symbol)
\section{Introduction}
Mass loss is the single most important process during the final evolution of low- and intermediate-mass stars on the asymptotic giant branch (AGB). Its existence and overall characteristics, e.g., magnitude, geometry and kinematics, are well established. However, much of its finer details, e.g., degree of asymmetry, non-homogeneity,  and temporal variation are essentially unknown. This is unfortunate since even a modest mass-loss-rate change by a factor of two will have a profound effect on the evolution of the star, its nucleosynthesis, and its return to the interstellar medium. 
The mass loss creates a circumstellar envelope (CSE) of gas and dust around
the star. The low temperature of the central star allows the formation of a wide variety of molecular species in its atmosphere, and the expanding gas is therefore mainly in molecular form.  The chemistry in the
CSE itself can be very rich and it depends on the C/O-ratio, the
thickness of the envelope, and the strength of the ambient ultraviolet
radiation field. 

Presented here is a short overview of various methods for estimating mass-loss rates  based on observations in the radio regime. Special attention will be given to CO line observations which has been considered, when supplemented by detailed radiative transfer modelling, as a very (perhaps the most) reliable AGB mass-loss-rate estimator. Uncertainties involved in the modelling procedure will be discussed.

\begin{figure}[!ht]
\plotone{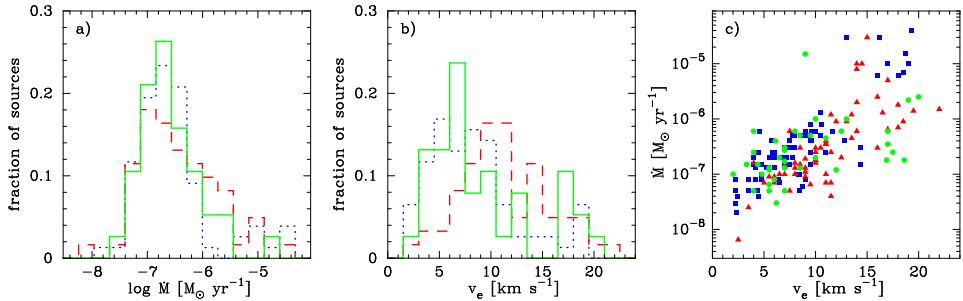}
\caption{{\bf a}) Mass-loss-rate distributions for the S-star (solid line; 40 stars; Ramstedt et al.\ 2006), M-star (dotted line; 77 stars; Olofsson et al.\ 2002; Gonz{\' a}lez Delgado et al.\ 2003), and carbon star (dashed line; 61 stars; Sch\"oier \& Olofsson 2001) samples. {\bf b}) Envelope gas expansion velocity distributions for the S-star (solid line), M-star (dotted line), and carbon star (dashed line) samples. 
{\bf c}) Derived mass-loss rates plotted against the gas expansion velocities for the S-star (dots), M-star (squares), and carbon star (triangles) samples.}
\nocite{Delgado03b}
\end{figure}

\section{Probing the mass-loss rate}
The discussion in this section will be limited to the more commonly used methods in deriving mass loss rates for AGB stars based on observations of spectral line emission in the radio regime. Although, radio continuum observations could in principle be used to drive dust mass-loss rates, and through the use of a dynamical model the gas mass-loss rates, they are mainly only included as part of more extensive multi-wavelength SED modelling which is usually better constrained from observations in the infra-red (and optical) part of the spectrum. In Sect.~4.2 a comparison between gas mass-loss rates derived from CO line emission and dust continuum emission is provided.

\subsection{HI 21\,cm line emission}
Hydrogen is the most abundant element in the material expelled from an AGB star. Molecular hydrogen (H$_2$) is difficult to observe since it has no allowed electric dipole rotational transitions and its (weaker) quadrupolar transitions lie in a spectral range in the infra-red that is not accessible from the ground. The atomic hydrogen (HI) line at 21\,cm could potentially be a useful probe of the gaseous envelopes around AGB stars. HI is also hard to ionize which would trace the CSE to large spatial scales. However, chemical models predict the gas in the wind to be atomic only when the central star has an effective temperature higher than about 2500\,K \citep[e.g.][]{Glassgold83}. HI can also be produced from photodissociation of H$_2$ by the interstellar radiation field at large radial distances from the star \citep{Gussie95}. There are some observational difficulties in that the emission is inherently weak and that HI emission from the interstellar medium provides an intense background from which the circumstellar contribution needs to be separated.

Mass-loss rates based on HI 21\,cm line observations are still very uncertain. This is partly because the HI abundance distribution may be complicated but there is also evidence that the AGB wind interacts with the ISM at large radial scales. Current HI mass-loss rate estimates show a correlation with those obtained from other methods, such as CO line observations, but the discrepancy for an individual source can be as large as an order of magnitude \citep{LeBertre04, Gerard06}. 

\subsection{CO line observations}
CO is usually the most abundant circumstellar molecule after H$_2$ and multi-transition CO line observations, when supplemented by a detailed excitation analysis, are commonly regarded as one of the most accurate methods for determining mass-loss rates on the AGB. The CO molecule is relatively easy  to excite even in low temperature low density regions and, like H$_2$, self-shileds against the interstellar uv-radiation field which combines to make it an excellent probe of the circumstellar medium. 

Several surveys of the mass-loss properties of AGB stars have been performed using CO line emission (e.g., Loup et al.\ 1993\nocite{Loup93};  Bieging \& Latter 1994\nocite{Bieging94}; Kahane \& Jura 1994\nocite{Kahane94}; Groenewegen et al.\ 1999\nocite{Groenewegen99}; Hiriart \& Kwan 2000\nocite{Hiriart00}; Sch\"oier \& Olofsson 2001\nocite{Schoeier01};  Olofsson et al.\ 2002\nocite{Olofsson02}; Ramstedt et al.\ 2006\nocite{Ramstedt06}; Teyssier et al.\  2006\nocite{Teyssier06}). It is found that the derived mass-loss-rate distribution has a median value of $\sim$\,2\,$\times$\,10$^{-7}$\,M$_{\odot}$\,yr$^{-1}$ with no obvious dependence on the photospheric C/O-ratio (see Fig.~1a). The mass-loss rates correlate well with the expansion velocity of the CSEs (Fig.~1b) and there are some indication that carbon stars generally have larger expansion velocities than S-stars and M-type AGB stars (Fig.~1c). The observed and derived properties of AGB CSEs and their correlation with stellar properties such as, e.g., luminosity and effective temperature lend support to the common concensus that these stellar winds are driven by radiation pressure on dust grains and that pulsation may play an important role \citep{Schoeier01,Bergeat05}.

Single-dish observations in the radio regime generally do not spatially resolve the CSEs around AGB stars. The first systematic survey of AGB molecular envelopes at high spatial resolution was performed by \citet{Neri98} using CO $J$\,$=$\,1\,$\rightarrow$\,0 line emission using the Plateu de Bure interferometer. They concluded that the CSEs generally have an overall spherical symmetry and expand isotropically at near constant expansion velocity. These results have further stimulated the continued use and development of the "standard model" described in Sect.~3.

\subsection{Maser emission}
Together with CO observations, maser line emission from molecules such as OH and H$_2$O (and to a lesser extent SiO) has commonly been used to derive mass-loss rates for AGB stars. This procedure is biased toward AGB stars in which the photospheric C/O-ratio is $<$\,1 which allows for the formation of high abundances of H$_2$O and its photodissociation product OH. The excitation conditions requires the maser emission to emanate from a region relatively close to the central star allowing for detailed studies of the inner part of the CSE where the mass-loss is initiated and the wind accelerated. Maser lines are usually very bright and can be used to study AGB stars at larger (extragalactic) distances \citep[e.g.,][]{VanLoon01}. 

Mass-loss rate estimates based on maser emission are inherently more uncertain than those obtained from CO measurements. 
Detailed radiative transfer modelling indicates that the maser strengths are very sensitive to changes not only in the chemistry, density, temperature, and velocity structures of the gas but also in the exact composition of the dust particles present in the wind \citep[e.g.,][]{Goldreich76, Babkovskaia06}. 
%Typically, at least for larger surveys, it is assumed that the measured intensity of the maser line scales with the far-infrared flux of the star which in turn is a measure of the mass loss rate \citep[e.g.,][]{}. 

\section{The standard model}
The numerical modelling of CSEs, where intricate interplays between physical
and chemical processes take place, is a challenge. This applies,
in particular, to the important inner regions of the CSEs, where very currently few
observational constraints are available and our knowledge is very
limited. Atomic and molecular line emission in the radio regime predominately probe the outer envelope regions where many of these complexities can be neglected and a fairly simple, yet reasonably realistic, CSE can be assumed. Here a brief summary of the assumptions used in the "standard model" often adopted when estimating mass-loss rates from CO line observations in the radio regime.  As an example a best fit model for the high mass-loss rate carbon star LP~And is shown in Fig.~2. In Sect.~4  some of the uncertainties involved in this modelling procedure will be discussed.

\subsection{Basic assumptions}
The most commonly adopted starting point for detailed modelling is to assume the CSEs to be spherically
symmetric, to be produced by a constant mass-loss rate, and to expand at a constant velocity. The density
structure, $\rho_{\mathrm{H}_2}$, (as a function of distance, $r$, from the central star) can
then be derived from the conservation of mass
\begin{equation}
\rho_{\mathrm{H}_2} = n_{\mathrm{H}_2} m_{\mathrm{H}_2} =
\frac{\dot{M}}{4\pi r^2 v_{\mathrm e}},
\end{equation}
where $\dot{M}$ is the hydrogen gas mass loss rate and $v_{\mathrm e}$ is
the gas expansion velocity taken from the observed line profiles.  
It is further assumed that the hydrogen is in molecular form in the region probed by the 
CO emission \citep{Glassgold83}. In the CO excitation analysis a CO/H$_2$ abundance ratio needs to be adopted. Typically the assumed ratio is in the range 2\,$\times$\,10$^{-4}$\,$-$\,1\,$\times$\,10$^{-3}$ with M-type AGB stars in the lower end and carbon stars in the upper end \citep[e.g.,][]{Cherchneff06}. 

\begin{figure}[!ht]
\plotone{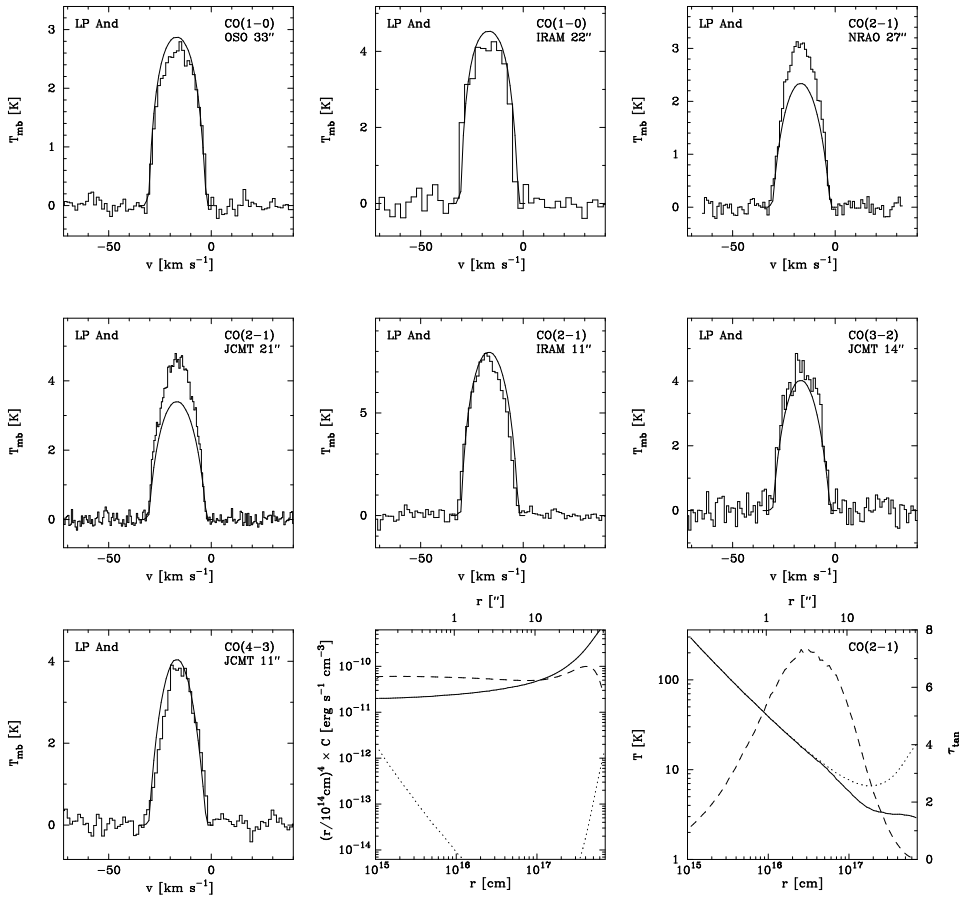}
  \caption{Multi-transition CO millimetre-wave line emission observations (histogram) of the carbon star LP And overlayed with model predicions (solid line). In the cooling panel the dotted and dashed lines are H$_2$ and CO line cooling, respectively. The solid line is adiabatic cooling due to expansion of the gas. In the temperature/optical depth panel the solid line is the excitation temperature of the CO $J$\,$=$\,2\,$\rightarrow$\,1 transition and the dotted line is the kinetic gas temperature. The dashed line is the tangential optical depth of the CO $J$\,$=$\,2\,$\rightarrow$\,1 transition at the line center (figure from Sch\"oier \& Olofsson 2001). }
\end{figure}

The CO molecules are generally not excited according to local thermodynamic equilibrium (LTE) in the regions of the envelope which contribute significantly to the observed emission. The common approach is instead to assume that statistical equilibrium is prevailing. The excitation analysis then requires a detailed knowledge of collisional rate coefficients between CO and H$_2$ molecules. Fortunately, CO-H$_2$ is a relatively well studied system (see summary in Sch\"oier et al.\ 2005b\nocite{Schoeier05a}) and calculated collisional rate coefficients currently exist for temperatures in the range 5\,$-$\,400\,K and involve energy levels up to $J$\,$=$\,29 and $J$\,$=$\,20 for collisions with para-H$_2$ and ortho-H$_2$, respectively. Extrapolations up to a collisional temperature of 2000\,K involving energy levels up to $J$\,$=$\,40, adequate for AGB CSEs, are provided by \cite{Schoeier05a} and available for download though the on-line database LAMDA ({\tt www.strw.leidenuniv.nl/$\sim$moldata}). In AGB CSEs an ortho-to-para ratio of 3 is usually adopted when weighting together collisional rate coefficients for CO in collisions with ortho- and para-H$_2$. Generally, and in particular for low mass-loss-rate objects, it is important that the excitation analysis also includes radiative excitation through the first vibrationally excited ($v$\,$=$\,1) state at 4.6\,$\mu$m (see Sect.~4.1). The models also include both a central source of radiation and the
cosmic microwave background radiation at 2.7\,K. The central radiation
emanates from the star itself, which may be approximated by a
blackbody. In some sources, where a prominent infrared-excess is observed,
fitting the spectral energy distribution with two blackbodies provide a
first approximation to the radiation field that the CSE is subjected to. Models that include an extended dust component in the CO excitation analysis may also be constructed \citep[e.g.,][]{Schoeier02b}.

The size of the CO envelope is an important parameter,
and the derived mass loss rate will depend on this 
(Sect.~4.1). If no radial brightness distributions exist the size of the circumstellar CO envelope can be estimated using the modelling presented in Mamon et al.\ (1988)\nocite{Mamon88}.  
It includes
photodissociation, taking into account the effects of dust-, self- and
H$_{2}$-shielding, and chemical exchange reactions. 
%\citet{Schoeier01} and \citet{Teyssier06}, using the interferometric observations by \citet{Neri98}, demonstrate that that the CO photodissociation calculation by Mamon et~al.\ gives reasonably accurate results for the majority of the sample stars.

\subsection{The energy balance equation}
Once the level populations are known the  kinetic gas temperature can be calculated in a self-consistent way
by solving the energy balance equation (e.g., Goldreich \& Scoville 1976),
\begin{equation}
\frac{dT}{dr} = (2-2\gamma)\frac{T}{r} + \frac{\gamma-1}
{n_{\mathrm H_2} k v_{\mathrm e}} (H-C),
\label{ebal}
\end{equation}
where $\gamma$ is the adiabatic index, $k$ the
Boltzmann  constant, $H$ the total heating rate per unit volume,
and $C$ the total cooling rate per unit volume.
The first term on the right hand side is
the cooling due to the adiabatic expansion of the gas.  Additional
cooling is provided by mainly molecular line emission from CO,
calculated from the derived level populations. 
In addition, HCN could
be an important coolant in the inner parts of the envelope of carbon stars and H$_2$O in the case of  M-type AGB stars.

The mechanism responsible for the observed mass loss is probably
 radiation pressure acting on
small dust grains, which in turn are coupled to the gas.
The radiation
pressure on the dust grains will give them a drift velocity,
$v_{\mathrm{dr}}$, relative to the gas (Gilman 1972\nocite{Gilman72};
Goldreich \& Scoville 1976\nocite{Goldreich76})
%; Kwan \& Hill 1977\nocite{Kwan77})
%
%\begin{equation}
%v_{\mathrm{dr}} = \left( \frac{L\,v_{\mathrm e}\,Q}{\dot{M} c}
%                  \right) ^{1/2},
%\label{v_dr}
%\end{equation}
%
%where $L$ is the luminosity, $Q$ the averaged momentum transfer efficiency,
%and $c$ the speed of light.
As a result of the dust-gas drift,
kinetic energy of the order of $\frac{1}{2} m_{\mathrm{H_2}} v^2_{\mathrm{dr}}$
will be transfered to the gas each time a particle collides with a dust grain.
This is assumed to provide the dominating  heating of the gas.
% (Goldreich \&
%Scoville 1976\nocite{Goldreich76}; Kwan \& Hill 1977\nocite{Kwan77}).
Additional heating is provided by, e.g., the photoelectric effect (Huggins et al.\ 1988\nocite{Huggins88}) 
that is important in the cool, tenuous, outer parts of CSEs
around (in particular) high mass loss rate stars.

When solving the energy balance equation several free parameters describing
the dust, i.e., via the dust-gas collisional heating, are introduced.
These are highly uncertain, but affect the derived line intensities.
For example, \citet{Schoeier01} 
%assume the $Q$-parameter, i.e., the efficiency of
%momentum transfer, to be equal to 0.03 
%(for details see Habing et al.\ 1994\nocite{Habing94}), 
%and 
define a new parameter that contains most of the these uncertainties,
\begin{equation}
h = \left(\frac{\Psi}{0.01}\right)
\left(\frac{2.0\,\mathrm{g\,cm}^{-3}}{\rho_\mathrm{d}}\right)
\left(\frac{0.05\,\mu\mbox{m}}{a_\mathrm{d}}\right),
\label{h}
\end{equation}
where $\Psi$ is the dust-to-gas mass-loss-rate ratio, $\rho_{\mathrm{d}}$ the density of a dust grain, $a_{\mathrm{d}}$ the radius of a dust grain. The $h$-parameter can be constrained if enough observational constraints are available (see Sect.~4.1).

\begin{figure}
\plotone{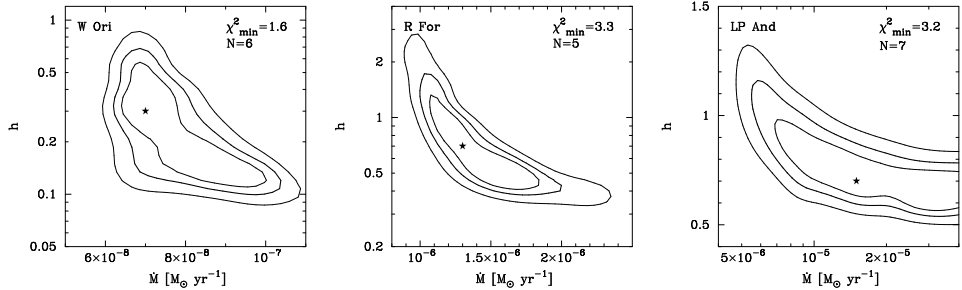}
  \caption[ ]{$\chi^2$-maps illustrating the sensitivity of the standard model to the adjustable parameters; the mass-loss rate ($\dot{M}$) and the $h$-parameter. W Ori is a low mass-loss-rate object, R For is of intermediate mass-loss rate and LP And is a high-mass-loss rate object. The thick contour marks the 1-$\sigma$ level of confidence (figure is adopted from Sch\"oier \& Olofsson 2001). }
\end{figure}

\section{Error analysis}
A full, detailed, error analysis of the estimated mass loss rates is not
possible due to the relatively large number of free parameters entering the
model. Instead,  some of the more important parameters may be varied in order to
illustrate the sensitivity of the model, and to be able to get a rough estimate
of the errors involved in the mass loss rate estimates. 

\subsection{Sensitivity tests}
%The results of these
%sensitivity tests are shown in Table~\ref{error} for our three example
%stars.  
Detailed sensitivity tests within the adopted circumstellar model (see Sect.~3) have been performed by several authors \citep{Kastner92,Groenewegen94,Schoeier01,Olofsson02}. 
In Fig.~3  chi-square contour plots
for three example stars (W Ori, R For, and LP And), produced by varying the mass loss rate and the
$h$-parameter, illustrating the accuracy in the determination
of these two adjustable parameters when other parameters are held fixed (except for the CO outer envelope radius which is allowed to vary according to Mamon et al.\ 1988\nocite{Mamon88}).

LP~And is a high mass-loss-rate object where the excitation of $^{12}$CO is dominated by collisions.
Consequently, the line intensities are very sensitive to the temperature
structure, i.e., the $h$-parameter. 
In the high mass loss rate regime the
line intensities are also insensitive to the adopted mass loss rate since an
increase in $\dot{M}$ leads to more cooling ($C$$\propto$$\dot{M}$
while $H$$\propto$$\dot{M}^{0.5}$), which compensates for the increase
of molecular density.  This "saturation"-effect of the line
intensities for high mass loss rates has been noted before in other
models where the cooling by CO is treated in a self-consistent manner
(Sahai 1990\nocite{Sahai90}; Kastner 1992\nocite{Kastner92}).
W~Ori, is a low mass loss rate object with low
optical depths in the lowest rotational transitions. In the
low mass-loss-rate regime the line intensities scale roughly linearly
with $\dot{M}$.  
As expected the radiation
emitted by the central star plays an important role in the excitation and the CO line intensities are less sensitive to the temperature structure ($h$-parameter).
R~For presents an intermediate case between LP~And and shares properties of both the low and the high mass-loss-rate stars.  An intermediate mass-loss-rate model is generally
more sensitive to the temperature structure than the radiation field,
thus resembling the high mass loss rate objects.  However, as in the
low mass loss rate regime the line intensities scale roughly linearly
with $\dot{M}$.  

In additional tests, \citet{Schoeier01} noted that
the assumed envelope size $r_{\mathrm p}$ does not significanly affect
the derived mass loss rate in the high mass loss rate regime since the 
density is too low to excite the CO
molecules effectively (the lines are sub-thermally excited) in the
cool outer parts of the envelope (cf.,  Fig.~2), i.e., the
emission is excitation limited.  In contrast, for low mass-loss-rate objects the size of the envelope is important
when determining the mass loss rate, i.e., the emission is
photodissociation limited (at least for the lowest transitions).  \citet{Schoeier01} made tests which
show that for typical envelope parameters the transition from photodissociation- to
excitation-limited emission lies at
$\sim$\,5\,$\times$10$^{-7}$\,M$_{\odot}$yr$^{-1}$ for the
$J$$=$$1$$\rightarrow$$0$ transition.  Higher $J$-lines sample hotter
and denser gas, located closer to the star, and are therefore less
sensitive to the choice of the envelope size.

In summary, it is concluded that within the adopted circumstellar model the
estimated mass loss rates are accurate to about $\pm$50\% (neglecting
errors introduced by the uncertain CO abundance and the distance
estimates) when good observational contraints are available,
but one should keep in mind that the causes of the
uncertainty varies with the mass loss rate.

\begin{figure}
\centerline{\psfig{figure=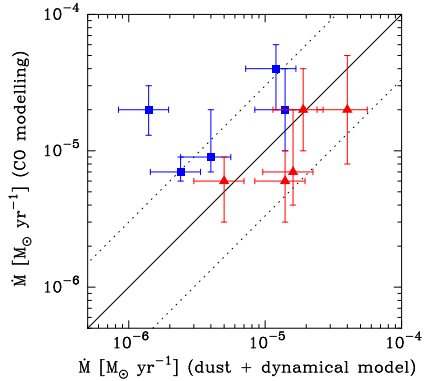,height=5cm,angle=-90}}
  \caption[ ]{Derived mass-loss rates from CO multi-line modelling and from SED modelling using a dust radiative transfer code coupled with a dynamical model. The sample consists of 5 M-type AGB stars (squares) and 5 carbon stars (triangles). The solid line marks the 1:1 correlation and the dotted lines a factor of three deviation (Ramstedt et al.\ 2007, in  prep.).}
\end{figure}

\subsection{Comparison between various authors and methods}
A rough way to estimate the errors involved is to compare with
the results obtained from other self-consistent models. \citet{Schoeier01} and recently \citet{Teyssier06} present comparisons of derived mass-loss rates from detailed modelling found in the literature for a large number of sources. They usually find a very good agreement, within 30\,\%, when adjustments for differences in adopted CO abundance and distance have been made. For an individual star, however, the discrepancy can be as large as a factor of three.

Dust radiative transfer modelling when coupled with a dynamical model of the wind can provide an independent estimate, to that of CO, of the gas mass-loss rate. In Fig.~4 derived gas mass-loss rates obtained from these two complementary methods are compared for a sample of intermediate- to high mass-loss-rate objects (Ramstedt  et al.\ 2007, in prep.). It is found that the two methods are generally consistent within a factor of three. Possible systematic effects here are the adopted CO fractional abundance and dust properties.

\begin{figure}
\plotone{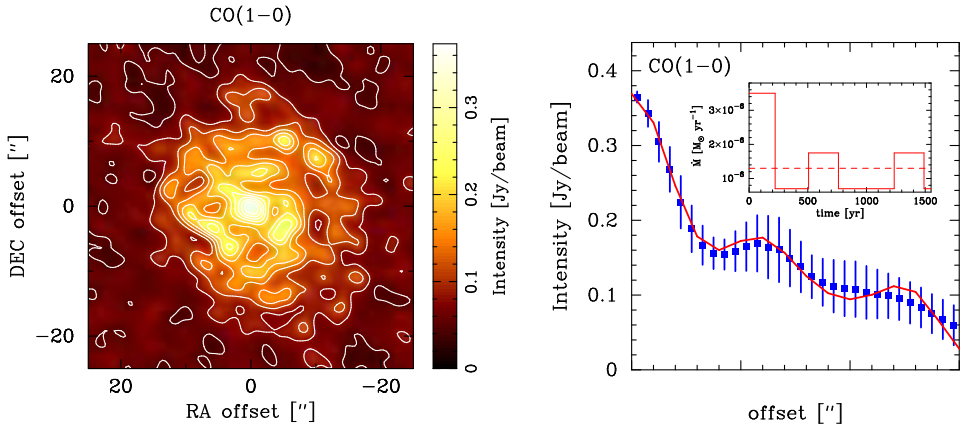}
       \caption[ ]{{\bf\em left --} Interferometric observations of CO line emission towards the M-type AGB star R Cas obtained by the PdBI (Castro-Carrizo et al.\ 2007, in prep.).  {\bf\em right --} The azimuthally averaged brightness distribution of the CO line emission around R Cas and the derived temporal modulation of the mass loss rate (Sch\"oier et al.\ 2007, in prep.). }
\end{figure}

\subsection{Departures from the standard model}
The sucessful modelling of the vast majority of AGB stars,
assuming a spherically symmetric CSE, suggests that pronounced axi- or non-symmetric mass
loss is not a common phenomenon among AGB stars.  However, there are cases where clear bipolar outflows have been detected such as the carbon star V~Hya \citep[e.g.,][]{Hirano04}. 

A small fraction ($\sim$\,10\%) of all carbon stars have detached molecular shells produced during a relatively short period of intense mass loss thought to be related to thermal pulses \citep[e.g.,][]{Olofsson96}. To be able to put constraints on the mass loss and its variation with time, high spatial resolution observations are needed. Radial brightness distributions also reveal smaller modulations of the mass-loss rate for a number of AGB stars (e.g., Sch\"oier \& Olofsson 2001\nocite{Schoeier01}; Teyssier et al.\ 2006\nocite{Teyssier06}; see also Fig.~5). Moreover, an  interacting wind scenario where the AGB wind sweeps up the ISM needs to be taken into account when interpreting mass-loss rate modulations \citep[e.g.,][]{LeBertre04,Schoeier05b} and the degree of clumpiness of the circumstellar medium needs to be fully investigated.

\section{Conclusions}
The mass loss rate determination for AGB stars depends crucially on a
number of assumptions in the CSE model. A reliable mass loss
rate determination requires, in addition to a detailed radiative
transfer analysis, good observational constraints in the form of
multi-transition observations and radial brightness distributions.
A standard CSE model, assuming a single smooth expanding wind produced by
a continuous mass loss, can explain multi-transition single-dish line observations in the majority of AGB stars and objects with clear deviations from the basic model assumptions can easily be identified. However, in order to investigate small deviations in terms of mass-loss rate modulations (typically lower than a factor of two), departure from spherical symmetry, or the degree of clumpiness, high-spatial-resolution interferometric observations are required. Such new observations are already becoming available and  reveal  the composition of AGB circumstellar envelopes are usually pathcy and clumpy to some degree also at larger radial scales from the central star. The next generation of radiative transfer codes, as well as dynamical and chemical models of circumstellar envelopes, need to take such effects into account.

\acknowledgements %%% Text of acknowledgements runs on after this command.
F.L.S. acknowledges financial support from the Swedish research council.

%%% THE BIBLIOGRAPHY
%%%
%%% CONSULT SECTION 3 OF "INSTRUCTIONS FOR AUTHORS" FOR HOW TO USE NATBIB.
%%% AUTHORS ARE ENCOURAGED TO USE EITHER THE "THEBIBLIOGRAPY" ENVIRONMENT
%%% BY UNCOMMENTING (DELETING THE "%" SYMBOL) THE COMMANDS BELOW, OR BY
%%% USING THE BIBTEX ENVIRONMENT. TO FIND OUT WHICH IS APPLICABLE TO YOUR
%%% CONTRIBUTION, CONSULT THE VOLUME EDITORS FOR YOUR PROCEEDINGS.
%%%

\bibliographystyle{aa}

\begin{thebibliography}
\expandafter\ifx\csname natexlab\endcsname\relax\def\natexlab#1{#1}\fi

\bibitem[{{Babkovskaia} \& {Poutanen}(2006)}]{Babkovskaia06}
{Babkovskaia}, N. \& {Poutanen}, J. 2006, \aap, 447, 949

\bibitem[{{Bergeat} \& {Chevallier}(2005)}]{Bergeat05}
{Bergeat}, J. \& {Chevallier}, L. 2005, \aap, 429, 235

\bibitem[{{Bieging} \& {Latter}(1994)}]{Bieging94}
{Bieging}, J.~H. \& {Latter}, W.~B. 1994, \apj, 422, 765

\bibitem[{{Cherchneff}(2006)}]{Cherchneff06}
{Cherchneff}, I. 2006, \aap, {in press}

\bibitem[{{G{\'e}rard} \& {Le Bertre}(2006)}]{Gerard06}
{G{\'e}rard}, E. \& {Le Bertre}, T. 2006, \aj, 132, 2566

\bibitem[{{Gilman}(1972)}]{Gilman72}
{Gilman}, R.~C. 1972, \apj, 178, 423

\bibitem[{{Glassgold} \& {Huggins}(1983)}]{Glassgold83}
{Glassgold}, A.~E. \& {Huggins}, P.~J. 1983, \mnras, 203, 517

\bibitem[{{Goldreich} \& {Scoville}(1976)}]{Goldreich76}
{Goldreich}, P. \& {Scoville}, N. 1976, \apj, 205, 144

\bibitem[{{Gonz{\' a}lez Delgado} {et~al.}(2003){Gonz{\' a}lez Delgado},
  {Olofsson}, {Kerschbaum}, {Sch{\" o}ier}, {Lindqvist}, \&
  {Groenewegen}}]{Delgado03b}
{Gonz{\' a}lez Delgado}, D., {Olofsson}, H., {Kerschbaum}, F., {et~al.} 2003,
  \aap, 411, 123

\bibitem[{{Groenewegen}(1994)}]{Groenewegen94}
{Groenewegen}, M.~A.~T. 1994, \aap, 290, 531

\bibitem[{{Groenewegen} {et~al.}(1999){Groenewegen}, {Baas}, {Blommaert},
  {Stehle}, {Josselin}, \& {Tilanus}}]{Groenewegen99}
{Groenewegen}, M.~A.~T., {Baas}, F., {Blommaert}, J.~A.~D.~L., {et~al.} 1999,
  \aaps, 140, 197

\bibitem[{{Gussie} {et~al.}(1995){Gussie}, {Taylor}, {Dewdney}, \&
  {Roger}}]{Gussie95}
{Gussie}, G.~T., {Taylor}, A.~R., {Dewdney}, P.~E., \& {Roger}, R.~S. 1995,
  \mnras, 273, 790

\bibitem[{{Hirano} {et~al.}(2004){Hirano}, {Shinnaga}, {Dinh-V-Trung}, {Fong},
  {Keto}, {Patel}, {Qi}, {Young}, {Zhang}, \& {Zhao}}]{Hirano04}
{Hirano}, N., {Shinnaga}, H., {Dinh-V-Trung}, {et~al.} 2004, \apjl, 616, L43

\bibitem[{{Hiriart} \& {Kwan}(2000)}]{Hiriart00}
{Hiriart}, D. \& {Kwan}, J. 2000, \apj, 532, 1006

\bibitem[{{Huggins} {et~al.}(1988){Huggins}, {Olofsson}, \&
  {Johansson}}]{Huggins88}
{Huggins}, P.~J., {Olofsson}, H., \& {Johansson}, L.~E.~B. 1988, \apj, 332,
  1009

\bibitem[{{Kahane} \& {Jura}(1994)}]{Kahane94}
{Kahane}, C. \& {Jura}, M. 1994, \aap, 290, 183

\bibitem[{{Kastner}(1992)}]{Kastner92}
{Kastner}, J.~H. 1992, \apj, 401, 337

\bibitem[{{Le Bertre} \& {G{\'e}rard}(2004)}]{LeBertre04}
{Le Bertre}, T. \& {G{\'e}rard}, E. 2004, \aap, 419, 549

\bibitem[{{Loup} {et~al.}(1993){Loup}, {Forveille}, {Omont}, \&
  {Paul}}]{Loup93}
{Loup}, C., {Forveille}, T., {Omont}, A., \& {Paul}, J.~F. 1993, \aaps, 99, 291

\bibitem[{{Mamon} {et~al.}(1988){Mamon}, {Glassgold}, \& {Huggins}}]{Mamon88}
{Mamon}, G.~A., {Glassgold}, A.~E., \& {Huggins}, P.~J. 1988, \apj, 328, 797

\bibitem[{{Neri} {et~al.}(1998){Neri}, {Kahane}, {Lucas}, {Bujarrabal}, \&
  {Loup}}]{Neri98}
{Neri}, R., {Kahane}, C., {Lucas}, R., {Bujarrabal}, V., \& {Loup}, C. 1998,
  \aaps, 130, 1

\bibitem[{{Olofsson} {et~al.}(1996){Olofsson}, {Bergman}, {Eriksson}, \&
  {Gustafsson}}]{Olofsson96}
{Olofsson}, H., {Bergman}, P., {Eriksson}, K., \& {Gustafsson}, B. 1996, \aap,
  311, 587

\bibitem[{{Olofsson} {et~al.}(2002){Olofsson}, {Gonz{\' a}lez Delgado},
  {Kerschbaum}, \& {Sch{\" o}ier}}]{Olofsson02}
{Olofsson}, H., {Gonz{\' a}lez Delgado}, D., {Kerschbaum}, F., \& {Sch{\"
  o}ier}, F.~L. 2002, \aap, 391, 1053

\bibitem[{{Ramstedt} {et~al.}(2006){Ramstedt}, {Sch{\"o}ier}, {Olofsson}, \&
  {Lundgren}}]{Ramstedt06}
{Ramstedt}, S., {Sch{\"o}ier}, F.~L., {Olofsson}, H., \& {Lundgren}, A.~A.
  2006, \aap, 454, L103

\bibitem[{{Sahai}(1990)}]{Sahai90}
{Sahai}, R. 1990, \apj, 362, 652

\bibitem[{{Sch{\" o}ier} {et~al.}(2005{\natexlab{a}}){Sch{\" o}ier},
  {Lindqvist}, \& {Olofsson}}]{Schoeier05b}
{Sch{\" o}ier}, F.~L., {Lindqvist}, M., \& {Olofsson}, H. 2005{\natexlab{a}},
  \aap, 436, 633

\bibitem[{{Sch{\" o}ier} \& {Olofsson}(2001)}]{Schoeier01}
{Sch{\" o}ier}, F.~L. \& {Olofsson}, H. 2001, \aap, 368, 969

\bibitem[{{Sch{\" o}ier} {et~al.}(2002){Sch{\" o}ier}, {Ryde}, \&
  {Olofsson}}]{Schoeier02b}
{Sch{\" o}ier}, F.~L., {Ryde}, N., \& {Olofsson}, H. 2002, \aap, 391, 577

\bibitem[{{Sch{\" o}ier} {et~al.}(2005{\natexlab{b}}){Sch{\" o}ier}, {van der
  Tak}, {van Dishoeck}, \& {Black}}]{Schoeier05a}
{Sch{\" o}ier}, F.~L., {van der Tak}, F.~F.~S., {van Dishoeck}, E.~F., \&
  {Black}, J.~H. 2005{\natexlab{b}}, \aap, 432, 369

\bibitem[{{Teyssier} {et~al.}(2006){Teyssier}, {Hernandez}, {Bujarrabal},
  {Yoshida}, \& {Phillips}}]{Teyssier06}
{Teyssier}, D., {Hernandez}, R., {Bujarrabal}, V., {Yoshida}, H., \&
  {Phillips}, T.~G. 2006, \aap, 450, 167

\bibitem[{{van Loon} {et~al.}(2001){van Loon}, {Zijlstra}, {Bujarrabal}, \&
  {Nyman}}]{VanLoon01}
{van Loon}, J.~T., {Zijlstra}, A.~A., {Bujarrabal}, V., \& {Nyman}, L.-{\AA}.
  2001, \aap, 368, 950

\end{thebibliography}

\end{document}